\documentclass{article}
\usepackage{spconf,amsmath,epsfig}
\usepackage{algorithm, algorithmic}
\usepackage{hyperref}
\usepackage{xcolor}
\usepackage{tikz}
\usepackage{subcaption}
\usepackage{amssymb}
\usepackage{graphics}
\usepackage{graphicx}
\usepackage{comment}
\usepackage{multirow}

\let\OLDthebibliography\thebibliography
\renewcommand\thebibliography[1]{
  \OLDthebibliography{#1}
  \setlength{\parskip}{0pt}
  \setlength{\itemsep}{0pt plus 0.3ex}
}

\begin{document}\sloppy
\def\x{{\mathbf x}}
\def\L{{\cal L}}

\title{Action-Constrained Reinforcement Learning for Frame-Level Bit Allocation in HEVC/H.265 through Frank-Wolfe Policy Optimization}
%
\name{Yung-Han Ho, Yun Liang, Chia-Hao Kao, Wen-Hsiao Peng}
\address{National Yang Ming Chiao Tung University, Taiwan}

\maketitle

\begin{abstract}
This paper presents a reinforcement learning (RL) framework that leverages Frank-Wolfe policy optimization to address frame-level bit allocation for HEVC/H.265. Most previous RL-based approaches adopt the single-critic design, which weights the rewards for distortion minimization and rate regularization by an empirically chosen hyper-parameter. More recently, the dual-critic design is proposed to update the actor network by alternating the rate and distortion critics. However, the convergence of training is not guaranteed. To address this issue, we introduce Neural Frank-Wolfe Policy Optimization (NFWPO) in formulating the frame-level bit allocation as an action-constrained RL problem. In this new framework, the rate critic serves to specify a feasible action set, and the distortion critic updates the actor network towards maximizing the reconstruction quality while conforming to the action constraint. Experimental results show that when trained to optimize the video multi-method assessment fusion (VMAF) metric, our NFWPO-based model outperforms both the single-critic and the dual-critic methods. It also demonstrates comparable rate-distortion performance to the 2-pass average bit rate control of x265.

\end{abstract}
\begin{keywords}
Frame-level bit allocation, rate control, action-constrained reinforcement learning
\end{keywords}
\section{Introduction}
\label{sec:introduction}


The task of frame-level bit allocation is to assign bits to every video frame in a group of pictures (GOP), aiming at minimizing the total distortion of a GOP subject to a rate constraint. In essence, it is a constrained optimization problem, with the inter-frame dependencies between the reference and non-reference frames incurring dependent decision-making.

Reinforcement learning (RL) is a promising technique for addressing dependent decision-making. Recently, some early attempts apply RL to the bit allocation problem for video coding. Chen \textit{et al.}~\cite{chen2018reinforcement} and Zhou \textit{et al.}~\cite{zhou2020rate} learn RL agents to determine quantization parameters (QP) for frame-level bit allocation. The former tackles the GOP coding with hierarchical bi-prediction, while the latter targets low-delay coding. Ren \textit{et al.}~\cite{ROIinVVC} extend the idea to region-of-interest (ROI)-based coding for gaming content. Fu \textit{et al.}~\cite{360HRL} adopt a similar approach to streaming applications. In common, these prior works all train the RL agent with a single reward function that usually mixes the distortion $r_D$ and the rate $r_R$ rewards by a fixed hyper-parameter $\lambda$, e.g. $r_D + \lambda r_R$. However, the choice of the hyper-parameter $\lambda$ is a non-trivial task. It affects crucially how the agent weighs the distortion minimization against the rate regularization. Using a fixed hyper-parameter may lead to an RL agent that does not generalize well to varied video sequences.

Different from the single-critic approaches ~\cite{chen2018reinforcement, zhou2020rate, ROIinVVC, 360HRL}, Ho \textit{et al.}\cite{ho2021dual} learn two separate critics, one for estimating the distortion $r_D$ reward and the other for the rate $r_R$ reward. They introduce a dual-critic learning algorithm that trains the RL agent by alternating the rate critic with the distortion critic according to how the RL agent behaves in encoding a GOP. If the agent violates the rate constraint, the rate critic is chosen to update the actor; otherwise, the distortion critic is applied to train the agent towards minimizing the distortion. Although the dual-critic scheme is found to generalize better on different types of video, the training convergence is not guaranteed.

In this paper, we propose an action-constrained RL framework though Neural Frank-Wolfe Policy Optimization (NFWPO). Similar to the dual-critic idea~\cite{ho2021dual}, our scheme includes a rate critic and a distortion critic. However, unlike~\cite{ho2021dual}, the rate critic is utilized to specify a state-dependent feasible set, i.e. an action space that meets the rate constraint. Next, we apply NFWPO~\cite{lin2021escaping} together with the distortion critic to identify a feasible action that works toward minimizing the distortion. It then follows that the chosen action is used as a target to guide the actor network. In particular, we choose the video multi-method assessment fusion (VMAF)~\cite{vmaf} as the distortion metric for two reasons. First,  VMAF can better reflect subjective video quality than mean-squared error (MSE). Second, we show that our RL framework is generic in that it can accommodate an intractable quality metric such as VMAF.   


Our main contributions are as follows: (1) this work presents a novel RL framework that incorporates the Frank-Wolfe policy optimization to address the frame-level bit allocation for HEVC/H.265; (2) it outperforms both the single-critic~\cite{chen2018reinforcement} and the dual-critic~\cite{ho2021dual} methods, showing comparable rate-distortion (R-D) results to the 2-pass ABR of x265. It is to be noted that our scheme performs bit allocation in one pass at test time. 

\section{Neural Frank-Wolfe Policy Optimization}
\label{sec:nfwpo}

Neural Frank-Wolfe Policy Optimization (NFWPO)~\cite{lin2021escaping} is an action-constrained reinforcement learning (RL) technique. Unlike the vanilla RL setup, the action-constrained RL requires the agent to maximize the reward-to-go $Q(s,a)$ subject to the feasible action constraints $\mathcal{C}(s)$:
\begin{equation}
\label{eq:NFWPO_objective}
\arg \max_{a\in \mathcal{C}(s)} Q(s,a),
\end{equation}
where the reward-to-go $Q(s,a)$ is the expected cumulative future rewards under the policy $\pi$ (i.e.~$Q(s,a) = E_{(s_t,a_t) \sim \pi}[\sum_{t=1}^\infty \gamma^{t-1} r(s_t,a_t)|_{s_1 = s, a_1 = a}]$). In this paper, the policy $\pi(s)$ is implemented by a deterministic, continuous actor network, which takes the state $s$ as input and produces a continuous action $a=\pi(s)$ as output. 

Some prior works \cite{dalal2018safe} deal with the action-constrained RL by adding a projection layer at the output of the actor network. The projection layer projects the action driven by the actor network onto the feasible set $\mathcal{C}(s)$ by 
\begin{equation}
\label{eq:feasible}
\prod\nolimits_{\mathcal{C}(s)}(a) = \arg \min_{y\in \mathcal{C}(s)} ||y-a||_2,
\end{equation}
where $a=\pi(s)$ is a pre-projection action and $\prod\nolimits_{\mathcal{C}(s)}(a)$ is the post-projection action. 
However, due to the projection layer, the zero-gradient issue may occur when updating the actor network by seeking the action which maximizes the reward-to-go $Q(s,\prod\nolimits_{\mathcal{C}(s)}(a))$ through the gradient ascent. One example of the projection layer is the ReLU operation, which may be employed to constrain the action to be a non-negative value. Apparently, it causes a zero gradient during back-propagation.

To circumvent this issue, NFWPO updates the actor network in three sequential steps. First, it identifies a feasible update direction $\bar{c}(s)$ according to \begin{equation}
\label{eq:c(s)}
\bar c(s)=\arg \max_{c\in \mathcal{C}(s)}\langle c,\nabla_{a} Q(s,a)|_{a=\prod\nolimits_{\mathcal{C}(s)}(\pi(s))}\rangle,
\end{equation} where the operator $\langle a,b\rangle$ takes the inner product of $a$ and $b$. 
Second, it evaluates a \textit{reference action} $\tilde{a}_s$ by
\begin{equation}
\label{eq:refa}
\tilde{a}_s=\prod\nolimits_{\mathcal{C}(s)}(\pi(s))+\alpha (\bar c(s)-\prod\nolimits_{\mathcal{C}(s)}(\pi(s)),
\end{equation}
where $\alpha$ is the learning rate of NFWPO.
Lastly, it learns the actor network $\pi(s)$ through gradient decent by minimizing the squared error between the reference action $\tilde{a}_s$ and $\pi(s)$: 
\begin{equation}
\label{eq:upt_actor}
\mathcal{L}_{NFWPO}=(\pi(s)-\tilde a_s)^2.
\end{equation}
It is worth noting that even though NFWPO still involves a projection layer, there is no zero-gradient issue during training because the projection layer does not participate in the back-propagation.



\section{Proposed Method}
\label{sec:pmethod}
\subsection{Problem Formulation}
\label{subsec:pform}
The objective of the frame-level bit allocation is to minimize the distortion (i.e.~maximize the VMAF score) of a GOP subject to a rate constraint. This is achieved by allocating available bits properly among video frames in a GOP through choosing their quantization parameters (QP). In symbols, we have
\begin{equation}
\label{eq:objective}
\arg \min_{\{QP_i\}}\sum_{i=1}^N D_i(QP_i)\text{ s.t.}\sum_{i=1}^N R_i(QP_i) = R_{GOP},
\end{equation}
where $QP_i$ indicates the QP for the i-th frame, $N$ denotes the GOP size, $D_i(QP_i)$ is the distortion (measured in VMAF) of frame $i$ encoded with $QP_i$, $R_i(QP_i)$ is the number of encoded bits of frame $i$, and $R_{GOP}$ is the GOP-level rate constraint. 

We address the problem in Eq.~\eqref{eq:objective} by learning an RL agent that is able to determine in sequential steps the QP of every video frame in a GOP, so that the resulting number of encoding bits approximates $R_{GOP}$ while maximizing the cumulative VMAF score. To this end, we introduce the NFWPO-based RL framework.

We begin by drawing an analogy between the NFWPO-based RL setup and the problem in Eq.~\eqref{eq:objective}. Consider a video frame $i$ whose $QP_i$ is to be determined. First, the minimization of the cumulative distortion over the entire GOP in Eq.~\eqref{eq:objective} is regarded as the maximization of the reward-to-go $Q$ in Eq.~\eqref{eq:NFWPO_objective} in determining $QP_i$, where $Q$ in our problem corresponds to the sum of VMAF scores from frame $i$ till the very last frame $N$ in the GOP. That is, the local QP decision for frame $i$ needs to ensure that the reward in the long run (respectively, the cumulative distortion till the very last frame) can be maximized (respectively, minimized). In addition, in determining $QP_i$, we wish to satisfy the rate constraint in Eq.~\eqref{eq:objective}. Obviously, not every possible $QP_i$ can meet the constraint. Therefore, $QP_i$ needs to be confined in a feasible set $\mathcal{C}(s_i)$, which is a function of the current coding context 
represented by a state signal $s_i$. How $\mathcal{C}(s_i)$ is specified is detailed in Section~\ref{subsec:RC-NFWPO}. With these in mind, we translate the problem in Eq.~\eqref{eq:objective} into 
\begin{equation}
\label{eq:reform_objective}
\arg \max_{QP_i\in \mathcal{C}(s_i)} Q (s_i,QP_i), \forall i,
\end{equation} which shares the same form as Eq.~\eqref{eq:NFWPO_objective}. We can thus learn an actor network for determining $QP_i$ based on the action-constrained RL setup (Section~\ref{sec:nfwpo}).   




\subsection{System Overview}
\label{subsec:oarch}
\begin{figure}[t!] 
\centering
\includegraphics[width=\linewidth]{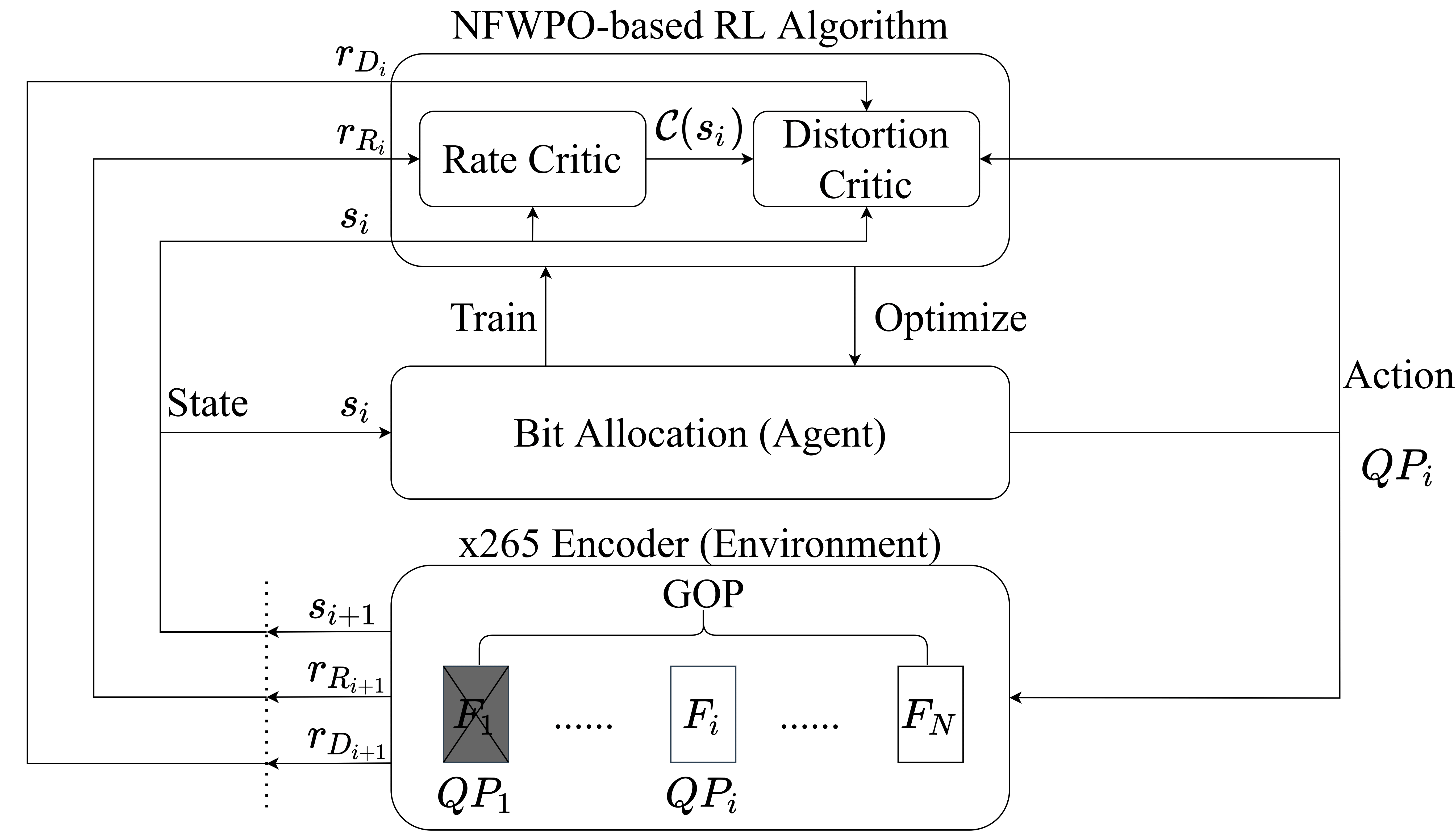}
\caption{The proposed NFWPO-based RL framework for frame-level bit allocation.}
\label{fig:RL}
\end{figure}

Fig.~\ref{fig:RL} presents an overview of our action-constrained RL setup. When encoding frame $i$, (1) a state $s_i$ is first evaluated (Section \ref{subsec:state}). (2) It is then taken by the neural network-based RL agent to output $QP_i$ as an action (Section \ref{subsec:action}). (3) With $QP_i$, frame $i$ is encoded and decoded by the specified codec (e.g. x265). Upon the completion of encoding frame $i$, our proposed NFWPO-based RL algorithm evaluates a distortion reward $r_{D_{i+1}}$ and a rate reward $r_{R_{i+1}}$ (Section \ref{subsec:rewards}). These steps are repeated iteratively until all the video frames in a GOP are encoded.

At training time, the agent interacts with the codec intensively by encoding every GOP as an episodic task. To predict the distortion and rate rewards, we resort to two neural network-based critics, the distortion critic $Q_D(s, QP)$ and the rate critic $Q_R(s,QP)$, both of which take the state signal $s$ and $QP$ as inputs. As will be seen, the rate critic $Q_R(s,QP)$, which predicts the rate reward-to-go (or the rate deviation from the bit budget $R_{GOP}$ at the end of encoding a GOP), allows us to specify a feasible set $\mathcal{C}(s_i)$ of $QP$ for the current coding frame. With $\mathcal{C}(s_i)$, we apply NFWPO~\cite{lin2021escaping} to train the agent, in order to maximize the distortion reward-to-go $Q_D(s,QP)$, which gives an estimate of the cumulative VMAF score. That is, we substitute $Q_D$ for $Q$ in Eq.~\eqref{eq:reform_objective}.

\vspace{-1em}

\subsection{State Signals}
\label{subsec:state}

The state signal provides informative information for the agent to make decisions. Inspired by~\cite{chen2018reinforcement}, we form our state signal with the following hand-crafted features. (1) The \textit{intra-frame feature} calculates the variance of pixel values for the current frame. Likewise, (2) the \textit{inter-frame feature} evaluates the mean of pixel values for the frame differences based on uni- and bi-prediction, where we adopt zero motion compensation as a compromise between performance and complexity.
(3) The \textit{average of the intra-frame features} averages the intra-frame features over the remaining frames that are not encoded yet. (4) The \textit{average of the inter-frame features} averages the inter-frame features over the remaining frames. Note that we refer to the original frames for computing frame differences whenever the coded reference frames are not available. (5) The \textit{number of remaining bits} in percentage terms evaluates $(R_{GOP} - R_{actual}) / R_{GOP}$, where $R_{GOP}$ and $R_{actual}$ are the GOP-level rate constraint and the number of bits encoded up to the current frame, respectively. 
(6) The \textit{number of remaining frames in the GOP} counts the frames that have not been encoded yet.
(7) The \textit{temporal identification} signals the level of the frame prediction hierarchy to which the current frame belongs.
(8) The \textit{rate constraint} of the GOP indicates the target bit rate for the current GOP.
(9) The \textit{base QP} specifies the default QP values for I-, B-, and b-frames, which are determined a priori irrespective of the input video. 

\vspace{-1.0em}
\subsection{Actions}
\label{subsec:action}

The action of our RL agent indicates the QP difference (known as the delta QP) from the base QP. That is, the final QP value (i.e. $QP_i$ in Fig.~\ref{fig:RL}) is the sum of the delta and the base QP's. In our implementation, the delta QP ranges from -5 to 5 irrespective of the frame type (I-, B-, or b-frame). The fact that the agent learns to decide the delta QP helps reduce the search space. Note that different frame types have different base QPs (see Section~\ref{subsec:esettings}).

\vspace{-1.0em}
\subsection{Rewards}
\label{subsec:rewards}
The ultimate goal of RL is to maximize the expected cumulative reward in the long run. In order for the agent to behave in a desired way, it is crucial to define proper rewards.

In this work, we specify two immediate rewards, the distortion reward $r_{D_i}$ and the rate reward $r_{R_i}$. The former $r_{D_i}$ is defined as
\begin{equation}
\label{eq:rewardd}
r_{D_i}=\text{VMAF}_i-\overline{\text{VMAF}}_i,
\end{equation}
where $\text{VMAF}_i$ and $\overline{\text{VMAF}}_i$ are the VMAF~\cite{vmaf} scores of frame $i$ when encoded with the QP values chosen by our learned agent and by the rate control algorithm of x265, respectively. In other words, $r_{D_i}$ characterizes the gain in VMAF achieved by our policy. This gain is intended to be maximized. 

The immediate rate reward, denoted by $r_{R_i}$, given to every video frame in a GOP is zero, except the last frame, the $r_{R_i}$ of which is given by 
\begin{equation}
\label{eq:Rreward}
r_{R_i}= 
\begin{cases}
    \frac{-|R_{GOP}-\sum_{t=1}^N R_t(QP_t)|}{R_{GOP}}, & \text{if } i=N; \\
    0, & \text{otherwise}. \\
\end{cases}
\end{equation}
The sum $\sum_{i=1}^N r_{R_i}$ represents the negative absolute deviation of the GOP bit rate from the target $R_{GOP}$ in percentage terms. 

With these immediate rewards, the distortion reward-to-go $Q_D(s,QP)$ and the rate reward-to-go $Q_R(s,QP)$ in Section~\ref{subsec:oarch} are evaluated as
\begin{equation}
\label{eq:D_reward-to-go}
Q_D(s_i,QP_i) = E_{(s_t,QP_t) \sim \pi}[\sum_{t=i}^N \gamma^{t-i} r_{D_t}]
\end{equation}
\begin{equation}
\label{eq:R_reward-to-go}
Q_R(s_i,QP_i) = E_{(s_t,QP_t) \sim \pi}[\sum_{t=i}^N \gamma^{t-i} r_{R_t}],
\end{equation}
where $\gamma=0.99$ is the discount factor. 
In particular, these reward-to-go functions are approximated by two separate networks, known as the distortion critic and the rate critic, respectively.


\begin{figure}[t]
\centering
\includegraphics[width=\linewidth]{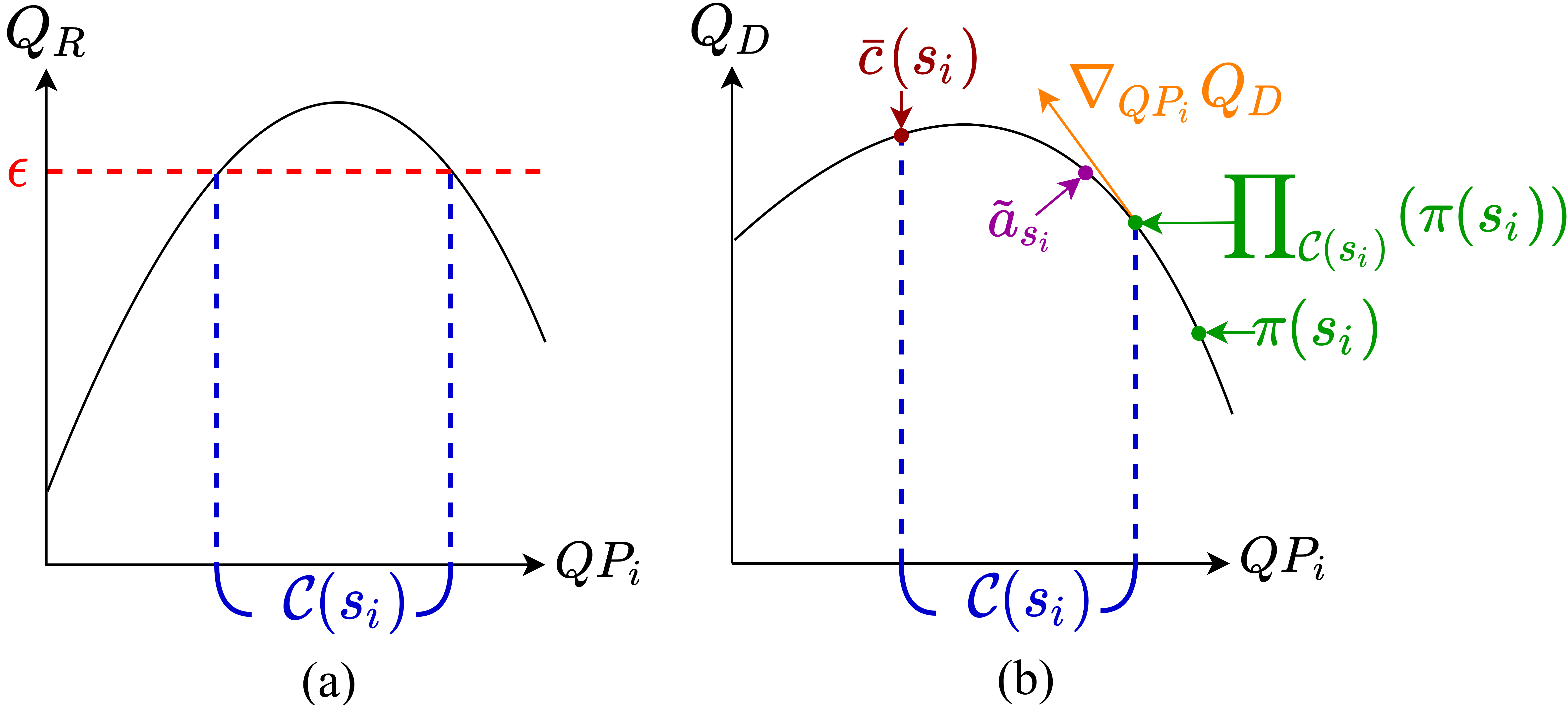}
\vspace*{-7mm}
\caption{Illustration of (a) the feasible set $\mathcal{C}(s)$, and (b) the reference action $\tilde{a}_s$.}
\label{fig:adapted NFWPO}
\vspace*{-5mm}
\end{figure}

\vspace*{-1em}
\subsection{NFWPO-based RL for Frame-level Bit Allocation}
\label{subsec:RC-NFWPO}


This section presents how we make use of the two critic networks to implement NFWPO. We begin with defining the feasible set $\mathcal{C}(s_i)$ in coding a video frame $i$. In order to meet the GOP-level rate constraint, the feasible set $\mathcal{C}(s_i)$ includes the QP values satisfying the requirement that the rate reward-to-go $Q_R$ is greater than or equal to a threshold $\epsilon$, given the current state $s_i$ (see Fig.~\ref{fig:adapted NFWPO} (a)):
\begin{equation}
\label{eq:rcfsetqr}
\mathcal{C}(s_i)=\{QP_i|Q_R(s_i,QP_i)\geq\epsilon\}.
\end{equation}
According to Eqs.~\eqref{eq:Rreward} and \eqref{eq:R_reward-to-go}, this implies that $\mathcal{C}(s_i)$ includes QP values that ensure the absolute rate deviation is upper bounded by the discounted $\epsilon$. In particular, we discretize these QP values by querying the rate critic $Q_R$ in discrete steps of 0.1 within the delta QP range (i.e. $QP_{i} = \{\text{base QP}-5, \text{base QP}-4.9, ..., \text{base QP}+5$\}). 

Given the feasible set $\mathcal{C}(s_i)$, we follow the process in Section \ref{sec:nfwpo} to generate the reference action $\tilde{a}_s$. Fig.~\ref{fig:adapted NFWPO} (b) illustrates the process, in which the actor output $\pi(s_i)$ is assumed to be outside of the feasible set. As shown, $\pi(s_i)$ will first be projected onto the feasible set to arrive at $\Pi_{\mathcal{C}(s_i)}(\pi(s_i))$. We then evaluate $\bar{c}(s_i)$ according to Eq.~\eqref{eq:c(s)}, where we use $Q_D(s,a)$ for $Q(s,a)$, and obtain $\tilde{a}_s$ based on Eq.~\eqref{eq:refa}. Finally, we update the actor network by Eq.~\eqref{eq:upt_actor}. 

Algorithm \ref{alg:A_NFWPO} details our proposed NFWPO-based RL algorithm. Similar to the DDPG algorithm, lines 5 to 12 roll out the learned policy with an exploration noise. The resulting transitions of states, actions, and rewards are stored in the replay buffer $R$ and are sampled to update the critics $Q_D$ and $Q_R$ in lines 15 and 17, respectively. Lines 19 to 22 correspond to NFWPO (Section \ref{sec:nfwpo}).

Our actor and critic networks have similar network architectures to~\cite{ho2021dual}, but take different state signals as inputs.

\begin{algorithm}[t]
\begin{footnotesize}
\caption{The proposed NFWPO-based RL algorithm}
\label{alg:A_NFWPO}
\begin{algorithmic}[1]
\STATE {Randomly initialize critics $Q_D(s,a|w_D), Q_R(s,a|w_R)$, and actor $\pi(s|\theta)$} with weights $w_D,w_R,$ and $\theta$
\STATE {Initialize target networks $Q_D'(s,a|w_D'), Q_R'(s,a|w_R')$, and $\pi'(s|\theta')$} with weights $w_D' \xleftarrow{} w_D,w_R' \xleftarrow{}w_R,$ and $\theta' \xleftarrow{}\theta$
\STATE {Initialize replay buffer $R$}
\STATE {\textbf{for} episode = $1$ to $M$ \textbf{do}}
\STATE {\quad Initialize a random noise process $\mathcal{N}$ for action exploration}
\STATE {\quad Evaluate initial state $s_1$}
\STATE {\quad \textbf{for} frame $i = 1$ to $N$ in a GOP \textbf{do}}
\STATE {\quad\quad Set $a_i = \pi(s_i|\theta)+\mathcal{N}_i$ }
\STATE {\quad\quad Encode frame $i$ with $QP=a_i$}
\STATE {\quad\quad Evaluate the immediate rewards $r_{D_i}$, $r_{R_i}$ and the new state $s_{i+1}$}
\STATE {\quad\quad Store transition $(s_i, a_i, r_{D_i}, r_{R_i}, s_{i+1})$ in $R$}
\STATE {\quad \textbf{end for}}
\STATE {\quad Sample $\mathcal{B}$ transitions $(s_{b}, a_{b}, r_{D_{b}}, r_{R_{b}}, s_{b+1})$ from $R$}
\STATE {\quad Set $y_{D_b} = r_{D_b}+\gamma Q_D'(s_{b+1},\pi'(s_{b+1}|\theta')|w_D')$ }
\STATE {\quad Update $Q_D$ by minimizing $\mathcal{L}=\frac{1}{\mathcal{B}}\Sigma_b{(y_{D_b}-Q_D(s_b,a_b|w_D))^2}$}
\STATE {\quad Set $y_{R_b} = r_{R_b}+\gamma Q_R'(s_{b+1},\pi'(s_{b+1}|\theta')|w_R')$ }
\STATE {\quad Update $Q_R$ by minimizing $\mathcal{L}=\frac{1}{\mathcal{B}}\Sigma_b{(y_{R_b}-Q_R(s_b,a_b|w_R))^2}$}
\STATE {\quad \textbf{for} each state s $\in \mathcal{B}$ \textbf{do}}
\STATE {\quad\quad Identify the feasible set $\mathcal{C}(s)$ by Eq.~\eqref{eq:rcfsetqr}}
\STATE {\quad\quad Obtain the update action $\bar{c}(s)$ by replacing $Q$ with $Q_D$ in Eq.~\eqref{eq:c(s)}}
\STATE {\quad\quad Obtain the reference action $\tilde{a}_s$ by Eq.~\eqref{eq:refa}}
\STATE {\quad\quad Update the actor network $\pi$ by Eq.~\eqref{eq:upt_actor}}
\STATE {\quad \textbf{end for}}
\STATE {\quad Every $c$ episodes, update the target networks: $Q_D'\xleftarrow{}Q_D, Q_R'\xleftarrow{}Q_R,$ and $\pi'\xleftarrow{}\pi$}
\STATE {\textbf{end for}}
\end{algorithmic}
\end{footnotesize}
\end{algorithm}
\section{Experimental Results}
\label{sec:eresults}
\vspace{-0.4em}

\subsection{Settings and Training Details}
\label{subsec:esettings}
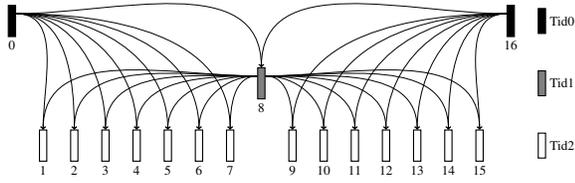
\begin{figure}
\centering
\resizebox{0.43\textwidth}{!}{%
\begin{tikzpicture}

\node[rectangle, minimum width = 0.00cm, minimum height =1cm, draw=black, line width=1pt,fill=black, draw ] (POC0) at (0,0){};
\coordinate [label=below:{\large 0}] (num0) at (POC0.south);

\node[rectangle, minimum width = 0.00cm, minimum height =1cm, draw=black, line width=1pt,fill=black!0, draw ] (POC1) at (1,-4){};
\coordinate [label=below:{\large 1}] (num1) at (POC1.south);

\node[rectangle, minimum width = 0.00cm, minimum height =1cm, draw=black, line width=1pt,fill=black!0, draw ] (POC2) at (2,-4){};
\coordinate [label=below:{\large 2}] (num2) at (POC2.south);

\node[rectangle, minimum width = 0.00cm, minimum height =1cm, draw=black, line width=1pt,fill=black!0, draw ] (POC3) at (3,-4){};
\coordinate [label=below:{\large 3}] (num3) at (POC3.south);

\node[rectangle, minimum width = 0.00cm, minimum height =1cm, draw=black, line width=1pt,fill=black!0, draw ] (POC4) at (4,-4){};
\coordinate [label=below:{\large 4}] (num4) at (POC4.south);

\node[rectangle, minimum width = 0.00cm, minimum height =1cm, draw=black, line width=1pt,fill=black!0, draw ] (POC5) at (5,-4){};
\coordinate [label=below:{\large 5}] (num5) at (POC5.south);

\node[rectangle, minimum width = 0.00cm, minimum height =1cm, draw=black, line width=1pt,fill=black!0, draw ] (POC6) at (6,-4){};
\coordinate [label=below:{\large 6}] (num6) at (POC6.south);

\node[rectangle, minimum width = 0.00cm, minimum height =1cm, draw=black, line width=1pt,fill=black!0, draw ] (POC7) at (7,-4){};
\coordinate [label=below:{\large 7}] (num7) at (POC7.south);

\node[rectangle, minimum width = 0.00cm, minimum height =1cm, draw=black, line width=1pt,fill=black!50, draw ] (POC8) at (8,-2){};
\coordinate [label=below:{\large 8}] (num8) at (POC8.south);

\node[rectangle, minimum width = 0.00cm, minimum height =1cm, draw=black, line width=1pt,fill=black!0, draw ] (POC9) at (9,-4){};
\coordinate [label=below:{\large 9}] (num9) at (POC9.south);

\node[rectangle, minimum width = 0.00cm, minimum height =1cm, draw=black, line width=1pt,fill=black!0, draw ] (POC10) at (10,-4){};
\coordinate [label=below:{\large 10}] (num10) at (POC10.south);

\node[rectangle, minimum width = 0.00cm, minimum height =1cm, draw=black, line width=1pt,fill=black!0, draw ] (POC11) at (11,-4){};
\coordinate [label=below:{\large 11}] (num11) at (POC11.south);

\node[rectangle, minimum width = 0.00cm, minimum height =1cm, draw=black, line width=1pt,fill=black!0, draw ] (POC12) at (12,-4){};
\coordinate [label=below:{\large 12}] (num12) at (POC12.south);

\node[rectangle, minimum width = 0.00cm, minimum height =1cm, draw=black, line width=1pt,fill=black!0, draw ] (POC13) at (13,-4){};
\coordinate [label=below:{\large 13}] (num13) at (POC13.south);

\node[rectangle, minimum width = 0.00cm, minimum height =1cm, draw=black, line width=1pt,fill=black!0, draw ] (POC14) at (14,-4){};
\coordinate [label=below:{\large 14}] (num14) at (POC14.south);

\node[rectangle, minimum width = 0.00cm, minimum height =1cm, draw=black, line width=1pt,fill=black!0, draw ] (POC15) at (15,-4){};
\coordinate [label=below:{\large 15}] (num15) at (POC15.south);

\node[rectangle, minimum width = 0.00cm, minimum height =1cm, draw=black, line width=1pt,fill=black, draw ] (POC16) at (16,0){};
\coordinate [label=below:{\large 16}] (num16) at (POC16.south);

\draw[->,black, thick] (POC0.60) to [out=0,in=90] (POC8.north);
\draw[->,black, thick] (POC0.60) to [out=0,in=90] (POC7.north);
\draw[->,black, thick] (POC0.60) to [out=0,in=90] (POC6.north);
\draw[->,black, thick] (POC0.60) to [out=0,in=90] (POC5.north);
\draw[->,black, thick] (POC0.60) to [out=0,in=90] (POC4.north);
\draw[->,black, thick] (POC0.60) to [out=0,in=90] (POC3.north);
\draw[->,black, thick] (POC0.60) to [out=0,in=90] (POC2.north);
\draw[->,black, thick] (POC0.60) to [out=0,in=90] (POC1.north);
\draw[->,black, thick] (POC16.120) to [out=180,in=90] (POC8.north);
\draw[->,black, thick] (POC16.120) to [out=180,in=90] (POC9.north);
\draw[->,black, thick] (POC16.120) to [out=180,in=90] (POC10.north);
\draw[->,black, thick] (POC16.120) to [out=180,in=90] (POC11.north);
\draw[->,black, thick] (POC16.120) to [out=180,in=90] (POC12.north);
\draw[->,black, thick] (POC16.120) to [out=180,in=90] (POC13.north);
\draw[->,black, thick] (POC16.120) to [out=180,in=90] (POC14.north);
\draw[->,black, thick] (POC16.120) to [out=180,in=90] (POC15.north);
\draw[->,black, thick] (POC8.60) to [out=0,in=90] (POC9.north);
\draw[->,black, thick] (POC8.60) to [out=0,in=90] (POC10.north);
\draw[->,black, thick] (POC8.60) to [out=0,in=90] (POC11.north);
\draw[->,black, thick] (POC8.60) to [out=0,in=90] (POC12.north);
\draw[->,black, thick] (POC8.60) to [out=0,in=90] (POC13.north);
\draw[->,black, thick] (POC8.60) to [out=0,in=90] (POC14.north);
\draw[->,black, thick] (POC8.60) to [out=0,in=90] (POC15.north);
\draw[->,black, thick] (POC8.120) to [out=180,in=90] (POC1.north);
\draw[->,black, thick] (POC8.120) to [out=180,in=90] (POC2.north);
\draw[->,black, thick] (POC8.120) to [out=180,in=90] (POC3.north);
\draw[->,black, thick] (POC8.120) to [out=180,in=90] (POC4.north);
\draw[->,black, thick] (POC8.120) to [out=180,in=90] (POC5.north);
\draw[->,black, thick] (POC8.120) to [out=180,in=90] (POC6.north);
\draw[->,black, thick] (POC8.120) to [out=180,in=90] (POC7.north);

\node[rectangle, minimum width = 0.025cm, minimum height =0.8cm, draw=black, line width=1pt,fill=black, draw ] (tid0) at (17,0){};
\coordinate [label=right:{\large Tid0}] (t0) at (tid0.east);

\node[rectangle, minimum width = 0.025cm, minimum height =0.8cm, draw=black, line width=1pt,fill=black!50, draw ] (tid1) at (17,-2){};
\coordinate [label=right:{\large Tid1}] (t1) at (tid1.east);

\node[rectangle, minimum width = 0.025cm, minimum height =0.8cm, draw=black, line width=1pt,fill=black!0, draw ] (tid2) at (17,-4){};
\coordinate [label=right:{\large Tid2}] (t2) at (tid2.east);

\end{tikzpicture}
}
\caption{The hierarchical B prediction structure. Frames 0 and 16 are I-frames, Frame 8 is B-frame, and the rest are b-frames.}
\label{fig:coding structure}
\vspace{-0.6em}
\end{figure}

We experiment with the proposed method on x265 using a hierarchical-B coding structure (GOP=16) as depicted in Fig. \ref{fig:coding structure}. To focus on the GOP-level bit allocation scheme, we follow the same GOP-level bit allocation as x265. To this end, we first encode every test sequence with fixed QP's 22, 27, 32, and 37. This establishes four target bit rates $R_s$ at the sequence level. Next, we enable the 2-pass ABR rate control mode of x265 (\textit{-{}-pass $2$ -{}-bitrate $R_s$ -{}-vbv-bufsize $2\times R_s$ -{}-vbv-maxrate $2\times R_s$}) in generating the anchor bitstreams, to meet these target bit rates. Finally, we observe the GOP-level bit rates of x265 and use them as our $R_{GOP}$. Since we use VMAF as the quality metric, to obtain the best perceptual quality, we turn off the \textit{-{}-tune} option in accordance with the x265 manual. It is to be noted that all the RL-based schemes (i.e. the single-critic method, the dual-critic method, and ours) are one-pass schemes. They do not make use of any information in the first encoding pass for bit allocation.



For training our model, the training dataset includes video sequences in UVG, MCL-JCV, and JCT-VC Class A. To expedite the training, all these sequences are resized to $512 \times 320$. Likewise, the test videos, which are Class B and Class C sequences in JCT-VC dataset, are resized to the same spatial resolution. The single-critic and dual-critic methods use the same dataset for training.

We determine the hyper-parameters as follows. The base QP are selected as $QP_l-3$, $QP_l-2$, and $QP_l+2$ for I-, B-, and b-frame, respectively, where $QP_l$ are 22, 27, 32, and 37. The learning rate $\alpha$ for NFWPO in Eq.~\eqref{eq:refa} is set to $0.1$, and the learning rate for the actor and critic networks is $0.001$. We use a discount factor $\gamma=0.99$ and the 3-step temporal difference learning to train our critics. 
The threshold $\epsilon$ of the feasible set is set to $-0.05$, allowing for a maximum rate deviation of $\pm5\%$. For a fair comparison, the same rate requirement applies to the single- and dual-critic methods.


\subsection{R-D Performance and Bit Rate Deviations}
\label{subsec:rd}
We assess the competing methods in terms of their R-D performance (in VMAF) and rate control accuracy. 

Table~\ref{tab:exp_BD_PSNR} presents BD-rate results, with the 2-pass ABR of x265 serving as anchor. 
From Table~\ref{tab:exp_BD_PSNR}, our scheme outperforms the single-critic~\cite{chen2018reinforcement} and the dual-critic~\cite{ho2021dual} methods, both of which exhibit $3\%-4\%$ rate inflation as compared to x265 (operated in ABR mode with two-pass encoding). Ours achieves comparable performance to x265, yet with one-pass encoding. One point to note here is that all three RL-based models, including ours, perform poorly on BQMall. When this out-liner is excluded, our 
scheme performs slightly better than x265, showing a 0.41\% average bit rate saving. 

To shed light on why our scheme does not work well on BQMall, we overfit our NFWPO-based model to the test sequences. In other words, the model is trained on the test sequences. The overfitting results presented in the right most column of Table~\ref{tab:exp_BD_PSNR} indicate that BQMall, on which our previous model performs poorly, benefit the most from overfitting. Moreover, the overfit model performs comparably to the non-overfit model on the other test sequences. This then implies that our NFWPO-based RL algorithm and the model capacity are less of a problem. The root cause of the poor performance on BQMall may be attributed to the less representative training data. 

Table~\ref{tab:exp_error} further presents the average bit-rate deviations from the GOP target $R_{GOP}$. The results are provided only for the highest and the lowest rate points due to the space limitation. In reporting the average rate deviation, any deviation from the $R_{GOP}$ within $\pm5\%$ is regarded as $0\%$ to account for our $\pm5\%$ tolerance margin.  
As seen from the table, our model shows fairly good rate control accuracy. The average rate deviations at high and low rates are around $2.1\%$ and $2.5\%$, respectively. In contrast, the rate deviations of the dual-critic method average $4.7\%$ and $4.9\%$ at high and low rates, respectively. The single-critic method has relatively inconsistent results: at high rates, the average rate deviation is as low as $1.2\%$, whereas at low rates, it is as high as $13.9\%$. This inconsistency in rate control accuracy between the low and high rates is attributed to the use of a fixed $\lambda$ for trading off the rate penalty against the distortion reward. 



\begin{table}[t]
\centering%
\caption{Comparison of BD-rates in terms of VMAF, with the 2-pass ABR of x265 serving as anchor. Ours* is the overfitting results of our NFWPO-based model.}
\label{tab:exp_BD_PSNR}
\setlength{\tabcolsep}{3mm}{
\scalebox{0.6}{
    \begin{tabular}{|c|c|c|c|c|}
    \hline
    \multirow{2}{*}{\textbf{Sequences}} &
    \multicolumn{4}{c|}{\textbf{BD-rate (\%)}}\\
    \cline{2-5}
      &Single&Dual&Ours&Ours*\\
    \hline
    BasketballDrill&3.14&7.88&0.09&\textbf{-0.03}\\
    \hline
    BasketballDrive&1.58&4.41&-0.01&\textbf{-0.45}\\
    \hline
    BQMall&6.66&3.38&5.01&\textbf{-0.36}\\
    \hline
    BQTerrace&8.07&0.20&0.98&\textbf{-2.57}\\
    \hline
    Cactus&2.00&0.81&\textbf{-1.39}&-0.35\\
    \hline
    Kimono&0.92&2.80&\textbf{-1.64}&-0.49\\
    \hline
    ParkScene&4.70&3.32&\textbf{-1.79}&0.23\\
    \hline
    PartyScene&5.19&4.29&1.43&\textbf{-1.61}\\
    \hline
    RaceHorses&1.71&1.02&\textbf{-0.93}&0.27\\
    \hline
    \hline
    Average (w/o BQMall)&3.42&3.09&-0.41&\textbf{-0.63}\\
    \hline
    Average&3.78&3.12&0.20&\textbf{-0.60}\\
    \hline

    \end{tabular}}
}
\end{table}

\begin{table}
\centering
\caption{Comparison of bit rate deviations from $R_{GOP}$.}
\label{tab:exp_error}
\scalebox{0.55}{
    \begin{tabular}{|c|c|c|c|c||c|c|c|c|}
    \hline
    \multirow{2}{*}{\textbf{Sequences}}&
    \multicolumn{1}{c|}{\textbf{Bit rate}}&
    \multicolumn{1}{c|}{\textbf{Single}}& \multicolumn{1}{c|}{\textbf{Dual}}& \multicolumn{1}{c||}{\textbf{Ours}}&
    \multicolumn{1}{c|}{\textbf{Bit rate}}&
    \multicolumn{1}{c|}{\textbf{Single}}& \multicolumn{1}{c|}{\textbf{Dual}}& \multicolumn{1}{c|}{\textbf{Ours}} \\
    \cline{2-9}
      &Kbps&Err. (\%)&Err. (\%)&Err. (\%)&Kbps&Err. (\%)&Err. (\%)&Err. (\%)\\
    \hline
    \multirow{1}{*}{BasketballDrill}&992&\textbf{0.0}&2.3&0.5
    &168&4.4&\textbf{1.2}&4.1\\
    \hline
    \multirow{1}{*}{BasketballDrive}&1093&\textbf{0.0}&3.7&0.9
    &171&1.3&1.1&\textbf{1.0}\\
    \hline
    \multirow{1}{*}{BQMall}&819&\textbf{0.0}&4.7&3.0
    &132&4.8&5.9&\textbf{3.6}\\
    \hline
    \multirow{1}{*}{BQTerrace}&691&\textbf{0.0}&4.2&1.1
    &93&28.2&10.9&\textbf{5.7}\\
    \hline
    \multirow{1}{*}{Cactus}&1108&7.4&3.3&\textbf{2.2}
    &158&6.0&1.6&\textbf{0.7}\\
    \hline
    \multirow{1}{*}{Kimono}&1022&1.8&6.3&\textbf{1.5}
    &139&16.8&9.9&\textbf{3.6}\\
    \hline
    \multirow{1}{*}{ParkScene}&881&\textbf{0.0}&2.0&\textbf{0.0}
    &103&50.1&3.6&\textbf{1.4}\\
    \hline
    \multirow{1}{*}{PartyScene}&1487&\textbf{0.4}&8.5&4.1
    &202&4.4&8.7&\textbf{1.9}\\
    \hline
    \multirow{1}{*}{RaceHorses}&1777&\textbf{1.3}&7.2&5.7
    &231&8.6&1.0&\textbf{0.3}\\
    \hline
    \hline
    \multirow{1}{*}{Average}&1097&\textbf{1.2}&4.7&2.1
    &155&13.9&4.9&\textbf{2.5}\\
    \hline

    \end{tabular}
}
\vspace{-1em}
\end{table}
\vspace{-0.5em}
\subsection{Effectiveness of Our NFWPO-based Model}
\label{subsec:effectivenessofNFWPO}
\begin{figure}[t]
\begin{center}
\begin{subfigure}{0.23\textwidth}
    \centering
    \includegraphics[width=\linewidth]{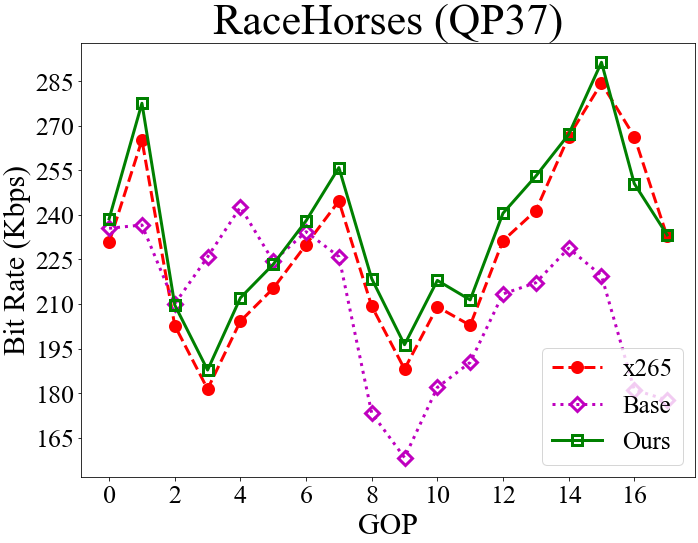}  
\end{subfigure}
\begin{subfigure}{0.23\textwidth}
    \centering
    \includegraphics[width=\linewidth]{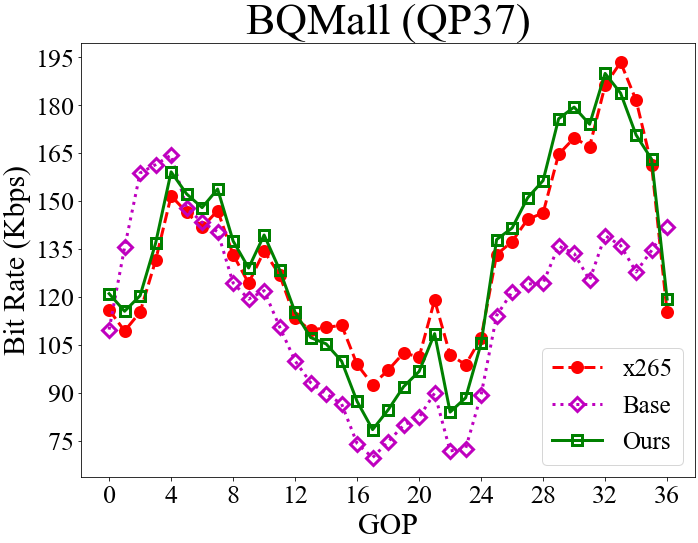}  
\end{subfigure}
\end{center}
\vspace{-2.0em}
\begin{center}
\begin{subfigure}{0.23\textwidth}
    \centering
    \includegraphics[width=\linewidth]{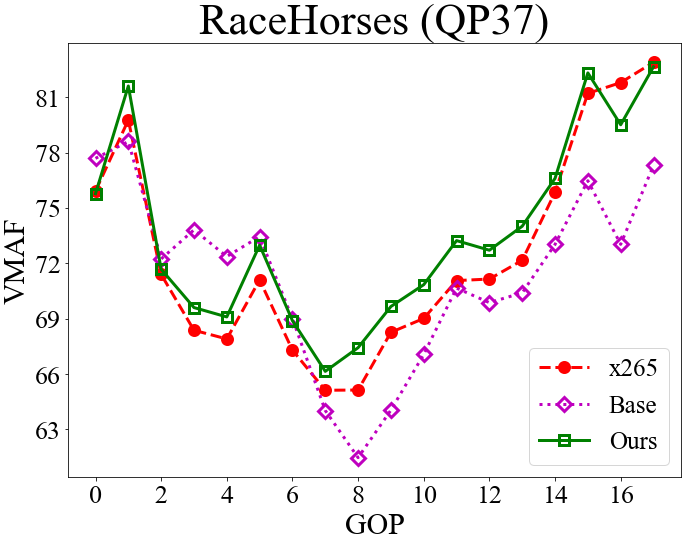}  
\end{subfigure}
\begin{subfigure}{0.23\textwidth}   
    \centering
    \includegraphics[width=\linewidth]{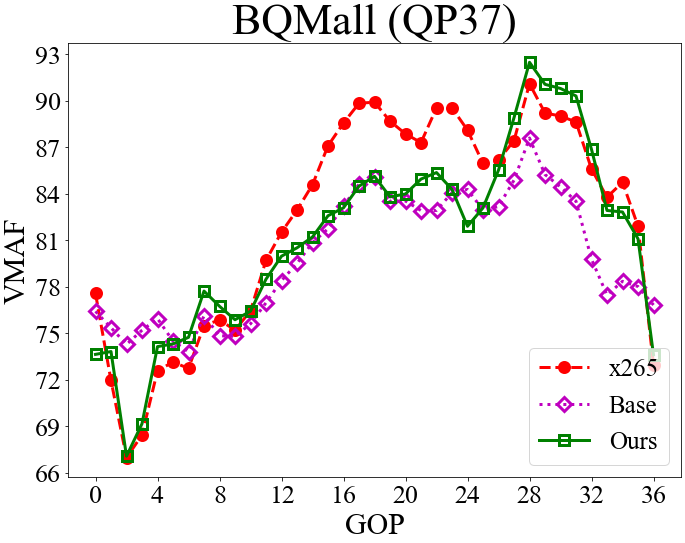}  
\end{subfigure}
\end{center}
\vspace{-2.0em}
\caption{Visualization of the GOP-level bit rates (top row) and VMAF scores (bottom row) for RaceHorses and BQMall.}
\label{fig:baserd}
\end{figure}

To understand the effectiveness of our rate control scheme, the top row of Fig.~\ref{fig:baserd} visualizes the GOP-level bit rates for RaceHorses and BQMall while the bottom row presents the corresponding VMAF scores. The objective here is to achieve better VMAF results when encoding every GOP at the same bit rate as x265 (Section~\ref{subsec:esettings}). As shown, applying the base QP alone, which corresponds to a fixed nested QP scheme, can hardly match the bit rates of x265. However, with our RL-based delta QP correction, the resulting bit rates match closely those of x265. Furthermore, our scheme shows similar or better VMAF scores than x265 in most GOPs, except for GOPs 12 to 24 in BQMall. This confirms the effectiveness of our RL training.


\begin{figure}[t]
\vspace{-1.0em}
\begin{center}
\begin{subfigure}{0.23\textwidth}
    \centering
    \includegraphics[width=\linewidth]{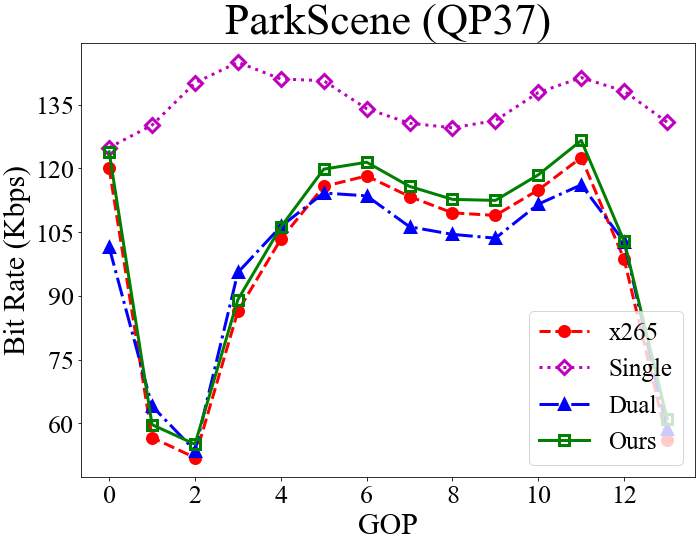}  
\end{subfigure}
\begin{subfigure}{0.23\textwidth}
    \centering
    \includegraphics[width=\linewidth]{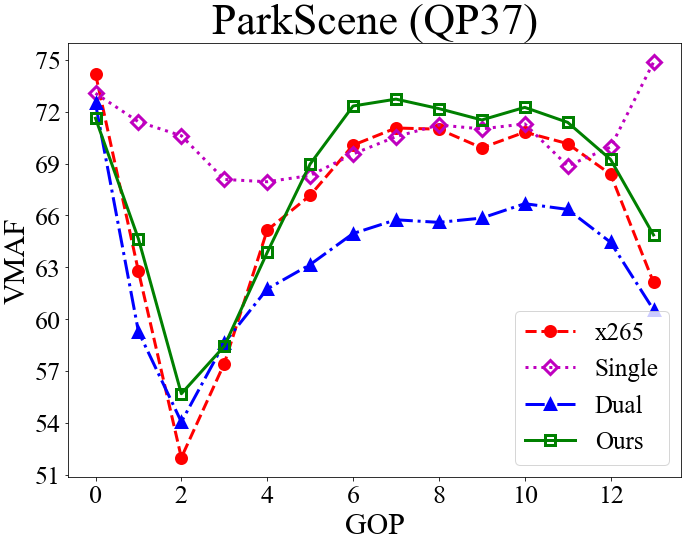}  
\end{subfigure}
\end{center}
\vspace{-2.0em}
\begin{center}
\begin{subfigure}{0.23\textwidth}
    \centering
    \includegraphics[width=\linewidth]{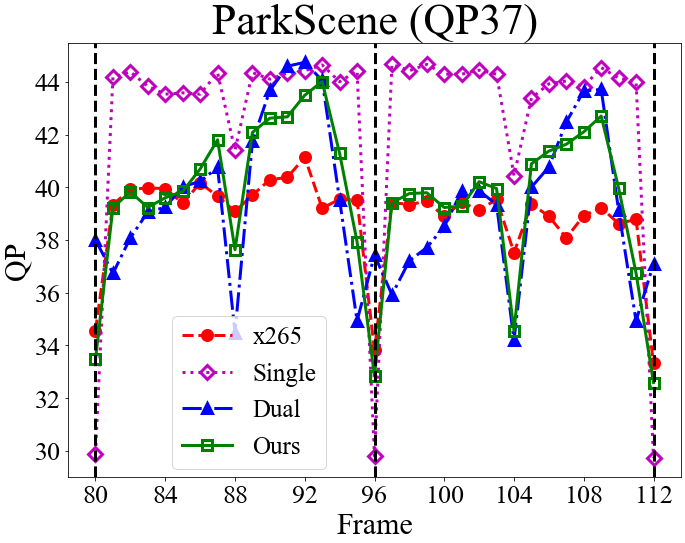}  
\end{subfigure}
\begin{subfigure}{0.23\textwidth}
    \centering
    \includegraphics[width=\linewidth]{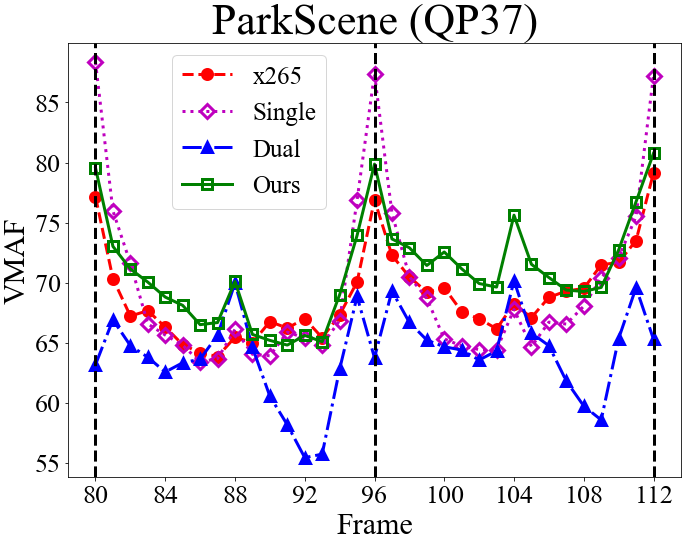}  
\end{subfigure}
\end{center}
\vspace{-2.0em}
\caption{Visualization of the GOP-level bit rates, the VMAF scores and the frame-level QP assignment within GOP 5 and 6 for ParkScene. Frames 80, 96, and 112 are I-frames.}
\vspace{-1em}
\label{fig:comparison_RL}
\end{figure}

\subsection{Rate Control Accuracy and QP Assignment}
\label{ComparisonRL}
Fig.~\ref{fig:comparison_RL} presents the GOP-level bit rates, the frame-level QP assignment, and the corresponding VMAF scores for the competing methods. Again, the objective is to match the GOP-level bit rate of x265 while maximizing the VMAF scores (Section~\ref{subsec:esettings}).


From the QP assignment, we see that the single-critic method tends to choose smaller QP values for I-frames, causing the resulting bit rate to exceed the target bit rate. The results presented correspond to the low-rate setting, where the single-critic method usually shows larger rate deviations (Section~\ref{subsec:rd} \& Table~\ref{tab:exp_error}). We attribute the reason to the use of a fixed $\lambda$ hyper-parameter in formulating a single reward function $r_D + \lambda r_R$. The varying dynamics between $r_D$ and $r_R$ at different bit rates and on different sequences may render the rate reward/constraint less prominent in some cases. In contrast, the dual-critic method has fairly precise rate control accuracy. It, however, shows lower VMAF scores because of choosing higher QP values for I-frames. The general observation is that at low rates, the dual-critic method tends to disregard the distortion in order to meet the stricter rate constraint. Different from these two methods, our model chooses similar QP values to x265 for I-frames. An intriguing finding is that it tries to improve the VMAF score by choosing relatively smaller QP values for the B-frame and the first group of b-frames (frames 1-7 in Fig.~\ref{fig:coding structure}) in a GOP. It then chooses larger QP values for the second group of b-frames (frames 9-15 in Fig.~\ref{fig:coding structure}) to meet the rate constraint. Another observation is that our NFWPO-based model is relatively more robust to varied coding conditions, showing good rate control accuracy at both low and high rates. 

\vspace{-0.5em}
\subsection{Subjective Quality Comparison}

\begin{figure}[t]
\centering
\includegraphics[width=1.05\linewidth]{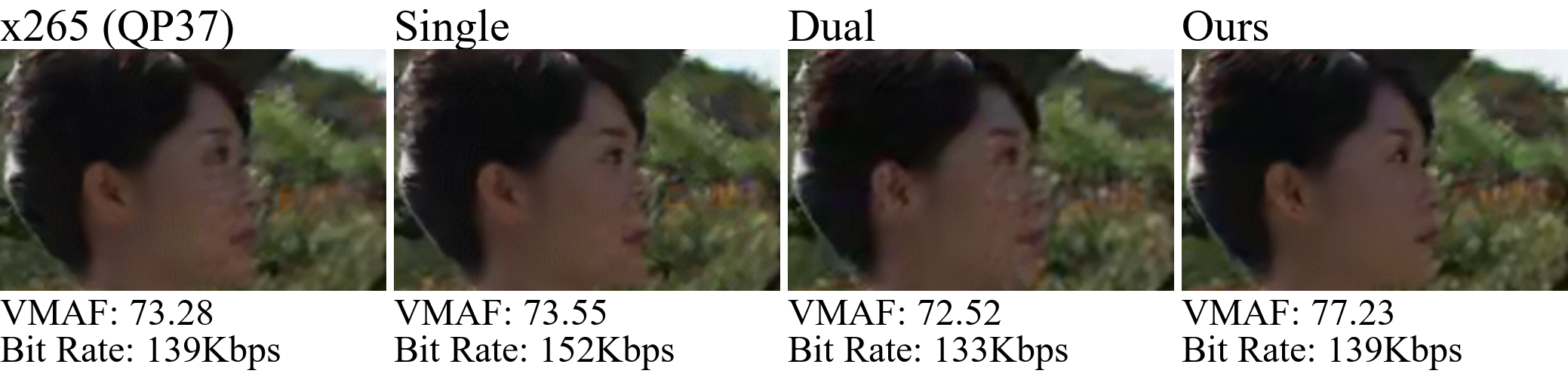}
\vspace*{-7mm}
\caption{Subjective quality comparison on Kimono.} 
\label{fig:subjectiv}
\vspace*{-5mm}
\end{figure}

Fig.~\ref{fig:subjectiv} presents a subjective quality comparison. We see that our proposed method preserves more texture details and shows sharper image quality. More subjective quality comparisons are provided in the supplementary document.



\section{Conclusion}
\label{sec:conclusion}
This paper presents a NFWPO-based RL framework for frame-level bit allocation in HEVC/H.265. It overcomes the empirical choice of the hyper-parameter in the single-critic method and the convergence issue in the dual-critic method. It outperforms both of these existing RL frameworks and shows comparable R-D results to the 2-pass average bit rate control of x265.
\vspace{-1em}

\vspace{-0.5em}
\bibliographystyle{IEEEbib}
\bibliography{references}

\end{document}